\documentclass[aps,prb,twocolumn,showpacs,superscriptaddress,nofootinbib]{revtex4-1}  
\usepackage{graphicx}  
\usepackage{dcolumn}   
\usepackage{bm}        
\usepackage{amsmath}
\usepackage{amsthm}
\usepackage{amssymb}
\usepackage{bbold} 
\usepackage{mathrsfs}
\usepackage{array}
\usepackage[usenames]{color}
\usepackage{soul} 
\usepackage{ulem} 
\begin{document}

\normalem

\title{Thermodynamic, magnetic and transport properties of the repulsive Hubbard model on the kagome lattice}
\author{Andressa R.~Medeiros-Silva}
\affiliation{Instituto de F\'isica, Universidade Federal do Rio de Janeiro, Rio de Janeiro, RJ 21941-972, Brazil}
\affiliation{Departamento de F\'isica, Universidade Federal do Piau\'i,
64049-550 Teresina PI, Brazil}
\author{Natanael C. Costa}
\affiliation{Instituto de F\'isica, Universidade Federal do Rio de Janeiro, Rio de Janeiro, RJ 21941-972, Brazil}
\author{Thereza Paiva} 
\affiliation{Instituto de F\'isica, Universidade Federal do Rio de Janeiro, Rio de Janeiro, RJ 21941-972, Brazil}

\begin{abstract}
Over the past decades, magnetic frustration has been under intense debate due to its unusual properties.
For instance, frustration in the kagome lattice suppresses long range spin correlations and it is expected to be a candidate for a spin liquid system.
Therefore, with the advent of experiments with ultra-cold atoms, the interest for frustrated geometries has increased.
Given this, in the present work we investigate the repulsive Hubbard model on the kagome lattice by unbiased quantum Monte Carlo simulations.
We examine its thermodynamic properties, as well as the magnetic and transport response of the system at finite temperatures and different values of the repulsive interaction.
From these results, we discuss the possible occurrence of adiabatic cooling, a quite important feature in ultra-cold systems, and the presence of a metal-to-insulator transition at a finite interaction strength.
Our findings may guide future experiments in ultra-cold fermionic atoms on the kagome lattice.
\end{abstract}


\maketitle

\section{Introduction}

\begin{figure}[t]
\includegraphics[scale=0.21]{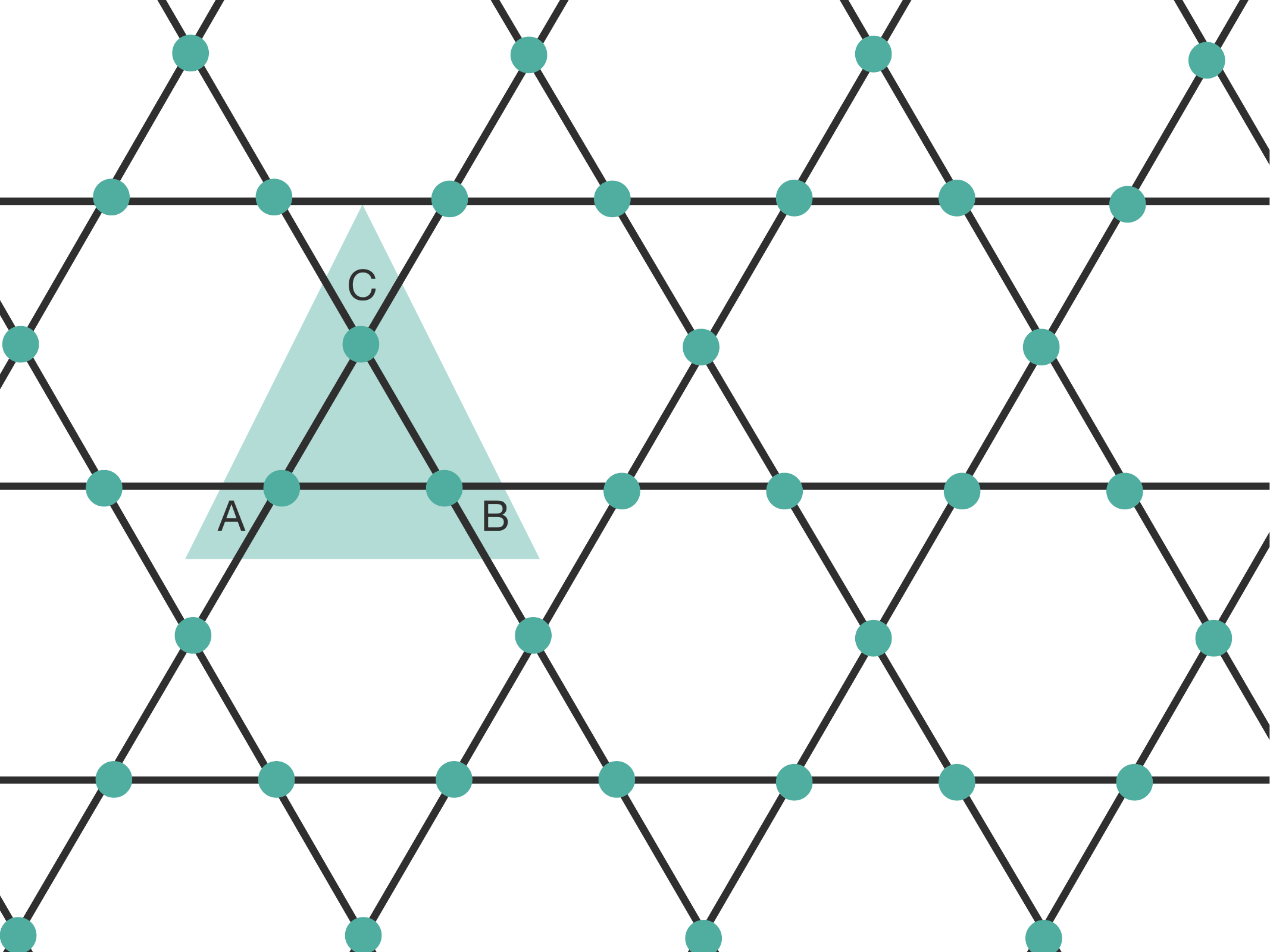} 
\caption{(Color online) The kagome lattice. The highlighted triangle describes the unit cell with its $A$, $B$ and $C$ basis sites.}
\label{fig:kagome}
\end{figure}

Magnetic frustration is a central issue in Condensed Matter Physics, with a great experimental effort being devoted, over the past decades, to understand its effects.
Indeed, the interest comes from the emergence of a myriad of correlated phases at low temperatures, due to the highly degenerate ground state of such systems.
Among the many geometries leading to frustration, the kagome lattice (see Fig.\,\ref{fig:kagome}) has gained much attention recently due to the possibility of the occurrence of a quantum spin liquid (QSL) state.
The experimental realization of this geometry is found, e.g., in herbertsmithite compounds\,\cite{Helton07,Helton10}, exhibiting strong evidences for the occurrence of QSL, although its nature is still under debate\,\cite{Han12,Fu15,Norman16}.
The electronic properties of the kagome lattice have also been investigated by the manipulation of atoms and molecules on given substrates\,\cite{Kempkes19,Sun20}, although the tuning of the interaction strength remains a challenge.
The advent of optical lattices has raised great expectations for unveiling fundamental properties of strongly correlated systems, in particular those with frustration.
Within this context, optical lattices for the kagome geometry were recently realized for bosonic atoms\,\cite{Jo12,Barter20,Leung20}, and one expects that fermionic ones could be realized in the near future.
In spite of the experimental effort, manipulating these many-body states is an arduous task, therefore providing information that could guide experiments is clearly in order.

From a theoretical point of view, the ground state properties of the Heisenberg model in the kagome lattice was extensively examined,
with the nature of its QSL phase (gapped or gapless) being a matter under debate\,\cite{Yan11,Iqbal13}.
However, this picture is less clear for the half-filled Hubbard model:
many studies, using different techniques, agree on the occurrence of a Mott transition from a finite value of interaction, although controversies on the critical value remain.
For instance, by using Variational Cluster Approximation (VCA), Yamada {\it et al}\,\cite{Yamada11} examined the occurrence of a metal-to-insulator (MIT) transition as the interaction strength increases, finding a critical point at $U_{c} /t \approx 5$, while Higa {\it et al}\,\cite{Higa16} suggested that such a transition should occur at $U_{c}/t = 6.8$.
Similarly, Ohashi {\it et al}\cite{Ohashi06} found this Mott transition around $U_{c}/t = 8.22$, using cellular dynamical mean field theory (CDMFT), while Variational Monte Carlo studies, conducted by Kuratani {\it et al}\,\cite{Kuratani07}, exhibits $U_{c}/t \approx 11$.
The lack of a consensus about $U_c$ on these studies may be due to the particularities of their implementations, i.e.~due to the way their biased input is added, and how it improves their ground states.

Unbiased methodologies are usually limited by technical issues, as the fermionic minus-sign problem for quantum Monte Carlo (QMC) approaches, or by dimensionality for DMRG.
Early attempts to perform finite temperature determinant QMC (DQMC) simulations were conducted by Bulut {\it et al}\,\cite{Bulut05}, without a clear evidence for a Mott transition.
Recently, by combining dynamical vertex approximation, dynamical mean-field theory, and DQMC, Kaufmann {\it et al}\,\cite{Kaufmann20} further examined how the magnetic correlation evolves in the kagome lattice, and proposed a critical point within a range of $U_{c}/t=[7,9]$.
On the other hand, DMRG results\,\cite{Sun21} provide evidence for two critical points, one for a translational symmetry broken insulator, and the other one to a QSL, at $U_{c1}/t \approx 5.4$ and $U_{c2}/t \approx 7.9$, respectively. 
Away from half-filling physical responses are even less clear, with the enhancement of unconventional pairing correlations\,\cite{Wen22}.

Despite these recent advances, much of the thermodynamic properties of the Hubbard model in the half-filled kagome lattice is unknown.
Knowing the many different energy scales of the system would be particularly important to cold atoms experiments.
In order to bridge this gap, in this work we perform a \textit{tour de force} using the DQMC method to examine the thermodynamic, magnetic and transport properties of such a frustrated system.
These analyses allow us to show the critical point within some accuracy, as well as unveiling hints about the ground state properties\,\cite{Aaram2020}.
The paper is organized as follows.
In Sec.\,\ref{sec:HQMC} we present the main features of the Hubbard Hamiltonian, and highlight the DQMC method together with the quantities of interest.
The results are presented in Secs.\,\ref{sec:results}, divided in subsections, in which we discuss the thermodynamic, magnetic, and transport properties, respectively.
Our main conclusions are then
summarized in Sec.\,\ref{sec:conc}.

\section{Model and Methodology}
\label{sec:HQMC}
Here we investigate fermions under a repulsive onsite interaction, namely the Hubbard model. Its symmetric Hamiltonian reads
\begin{align}\label{Eq:Hamil}
\nonumber \mathcal{H} = & -t\sum_{\substack{\langle \textbf{i},\textbf{j} \rangle},\sigma} \big( c_{\textbf{i} \sigma}^{\dagger}c_{\textbf{j} \sigma}+ {\rm H.c.} \big) - \mu \sum_{\substack{\textbf{i}}, \sigma} n_{\textbf{i},\sigma}
\\  & + U   \sum_{\substack{\textbf{i}}} \big(n_{\textbf{i},\uparrow} - 1/2 \big) \big(n_{\textbf{i},\downarrow} - 1/2\big),
\end{align}
where the sums run over sites of the kagome lattice, with $\langle \mathbf{i}, \mathbf{j} \rangle$ denoting nearest-neighbour sites under periodic boundary conditions.
In Eq.\,\eqref{Eq:Hamil}, we use the second quantization formalism, with $c^{\dagger}_{\mathbf{i} \sigma}$ ($c^{\phantom{\dagger}}_{\mathbf{i} \sigma}$) being creation (annihilation) operators of electrons on a given site $\mathbf{i}$, and spin $\sigma$, while $n_{\mathbf{i}\sigma} \equiv c^{\dagger}_{\mathbf{i} \sigma} c_{\mathbf{i} \sigma}$ are number operators.
The first two terms on the right hand side of the Hamiltonian correspond to the hopping of fermions, and the chemical potential $\mu$, respectively, with the latter determining the filling of the bands.
The third term describes the local repulsive interaction between fermions, with coupling strength \textit{U}.
Hereafter, we define the lattice constant as unity, and the hopping integral $t$ as the energy scale.

\begin{figure}[t]
\includegraphics[scale=0.28]{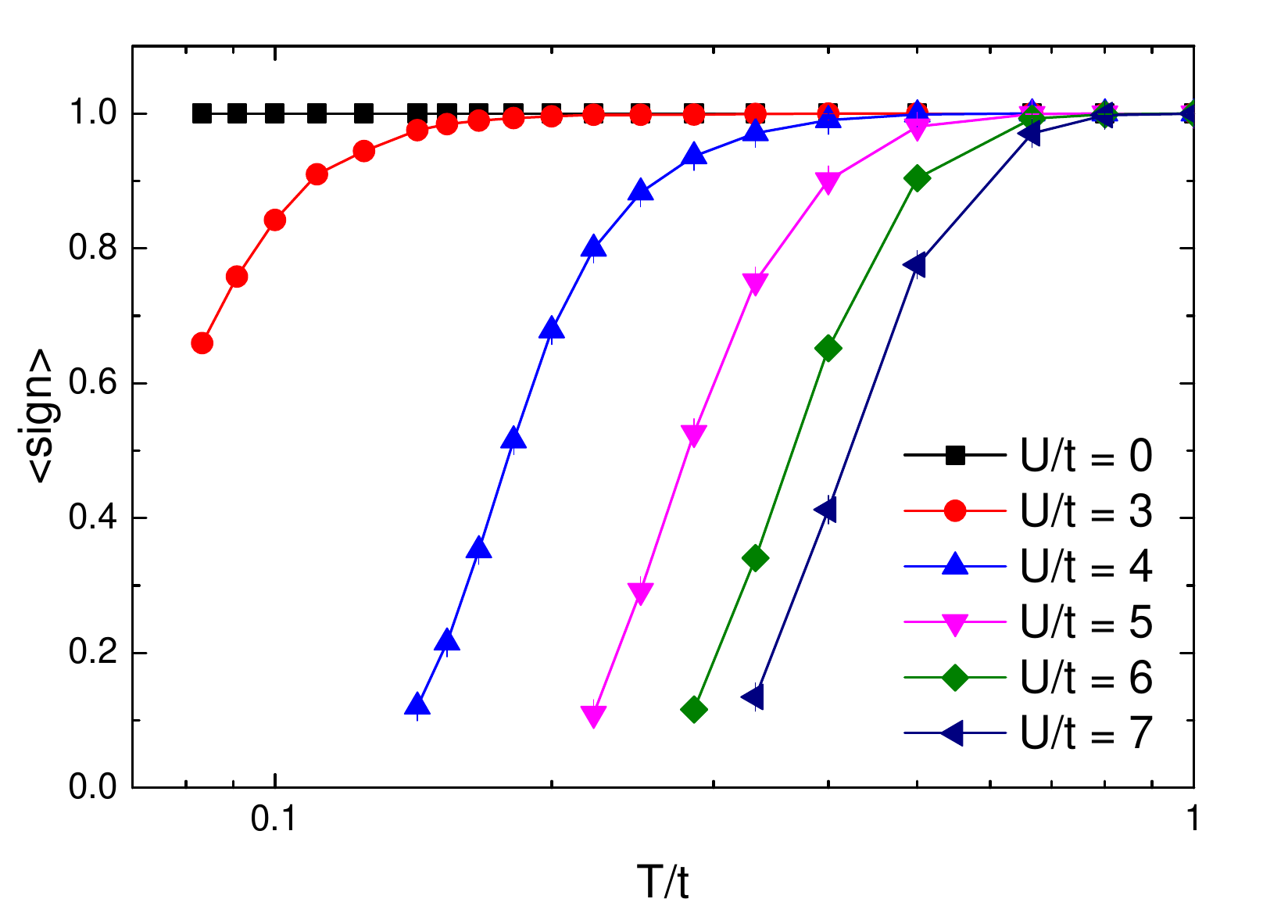} 
\caption{(Color online) The average fermionic sign as a function of the temperature, for different values of $U/t$. When not shown, error bars are smaller than symbol size.}
\label{fig:sign}
\end{figure}

We investigate the thermodynamic properties of Eq.\,\eqref{Eq:Hamil} on the half-filled kagome lattice by performing determinant quantum Monte Carlo (DQMC) simulations~\cite{Blankenbecler81,Hirsch83,Hirsch85,White89}.
The DQMC method is an unbiased numerical approach which maps a many-particle interacting fermionic system into a single-particle (quadratic form) one, with the aid of bosonic auxiliary fields.
In summary, the method separates the exponentials of the one-body and two-body terms, $\hat{\mathcal K}$ and $\hat{\mathcal P}$, respectively, in the partition function by performing a Trotter-Suzuki decomposition, i.e.
${\cal Z} = {\rm Tr}\, e^{-\beta \hat {\mathcal H} }
= {\rm Tr}\, \big[ \big(
e^{-\Delta\tau ( \hat{\mathcal K} + \hat{\mathcal P})} \big)^{L_{\tau}} \big] 
\approx {\rm Tr}\, \big[ 
e^{-\Delta\tau \hat{\mathcal K}} 
e^{-\Delta\tau \hat{\mathcal P}} 
e^{-\Delta\tau \hat{\mathcal K}} 
e^{-\Delta\tau \hat{\mathcal P}} \cdots \big]$.
Here, $L_{\tau}=\beta/ \Delta \tau$ is the size of the imaginary-time coordinate, corresponding to the number of incremental time evolution operators, with the inverse temperature $\beta \equiv 1/ (k_B T)$ where $k_B$ is the Boltzmann constant.
Such a decomposition has an error proportional to $(\Delta \tau)^2$, being exact in the limit $\Delta \tau \to 0 $.
In this work, we choose $t \Delta \tau \leq 0.05$, so that the error from the Trotter-Suzuki decomposition is negligible compared to that from the Monte Carlo sampling.

Proceeding, we perform a discrete Hubbard-Stratonovich (HS) transformation to rewrite quartic operators into quadratic (single-particle) ones, but at the cost of introducing auxiliary fields $s(\mathbf{i},\tau)$ on both real and imaginary-time coordinates, which are sampled by the regular Monte Carlo techniques.
Finally, from the Green's function, and, by using Wick contractions, all the higher-order correlation functions may be obtained.
More details about this methodology are discussed in Refs.\,\onlinecite{Santos03,assaad02,gubernatis16,Becca17} and references therein.

\section{Results}
\label{sec:results}

\subsection{Sign and the half filling}

\begin{figure}[t]
\includegraphics[scale=0.28]{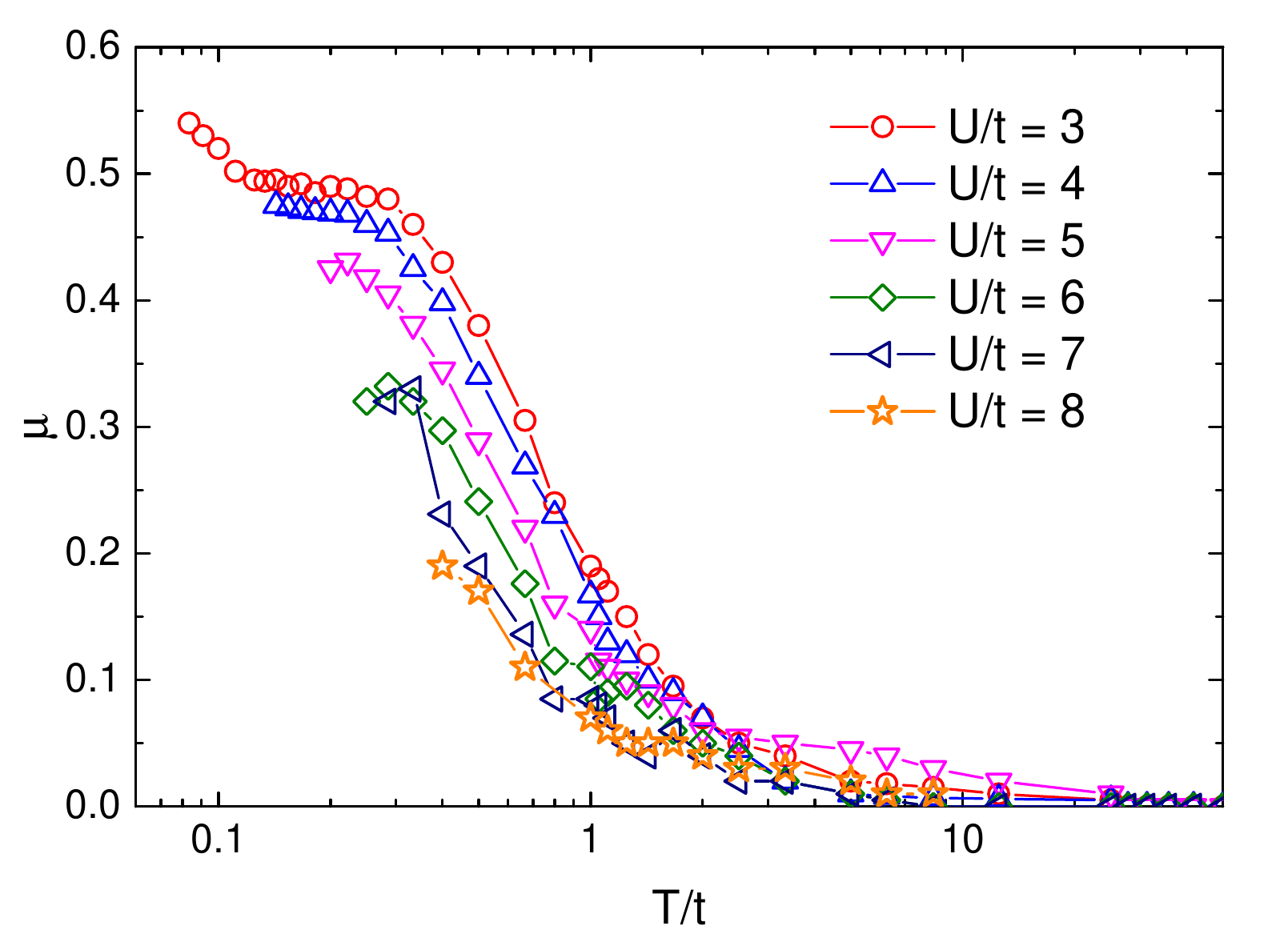} 
\caption{(Color online) The chemical potential that leads to the half-filling of the system as a function of the temperature, and for different values of interaction strengths.}
\label{fig:mu}
\end{figure}

Although being an unbiased methodology, the DQMC suffers with the infamous minus-sign problem, leading to noisy averages~\cite{Loh90,Troyer05}.
This problem does not exist at half-filled systems with particle-hole symmetry (PHS), such as bipartite lattices, as the square and the honeycomb ones.
However, the kagome lattice is non-bipartite, and there is no PHS for any filling, which, in turn, may lead to a severe sign problem depending on the system size, temperature scale, and interaction strength.
To further illustrate it, Fig.\,\ref{fig:sign} shows the average sign as a function of temperature $T/t$ at half-filling on the kagome lattice, for the different $U/t$ values, and for fixed a linear size $L=6$, i.e.~$N=6\times6\times3$ sites.
Notice that the sign decreases as the temperature is lowered, and it is strongly suppressed as $U/t$ increases (this behavior is more accentuated for larger system sizes).
Therefore, the following results are obtained for $L=6$, keeping $\langle {\rm sign} \rangle \gtrsim 0.05$, which, for some cases, demand simulations up to $5\times10^6$ Monte Carlo sweeps for measurements.
Similarly, unless otherwise indicated, the following results for the noninteracting case ($U/t=0$) are obtained in the thermodynamic limit.

Another important feature of the kagome lattice is that, due to the absence of the PHS, half-filling is not at $\mu=0$. Therefore, one needs to vary the chemical potential in order to find which $\mu$ leads to $\langle n \rangle =1$.
Figure \ref{fig:mu} shows how the chemical potential leading to a half-filled band at different $U/t$ changes with temperature.

\begin{figure}[t]
\includegraphics[scale=0.23]{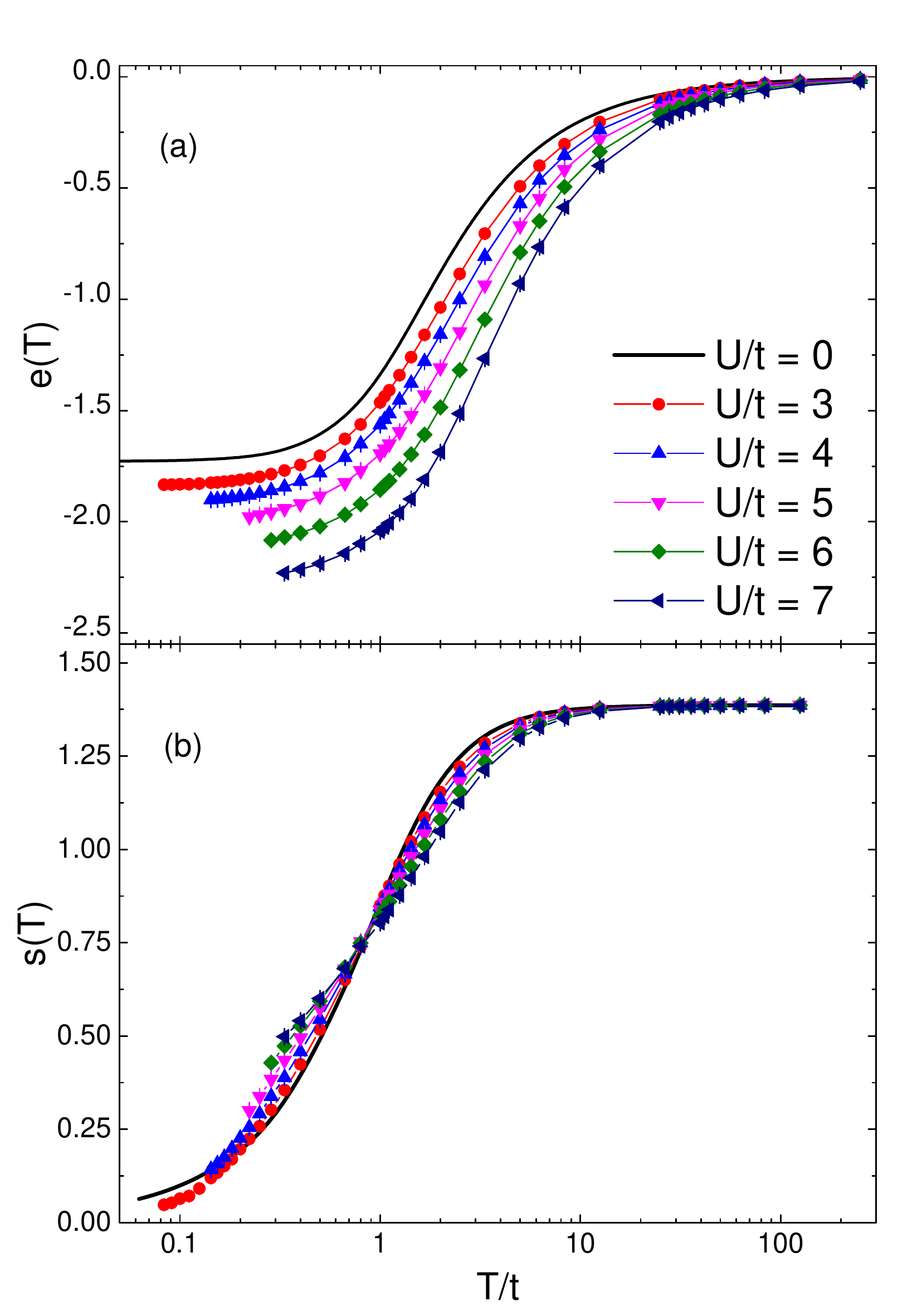} 
\caption{(Color online) The (a) total energy, and (b) entropy as functions of the temperature, for different values of $U/t$. When not shown, error bars are smaller than symbol size.}
\label{fig:enerentr}
\end{figure}

\subsection{Thermodynamic properties}
We start our analysis by discussing the thermodynamic properties of the system. 
First, we investigate the internal energy density,
\begin{equation}
\label{energia}
    e(\beta, U) = \frac{1}{N} \langle \mathcal{H} \rangle,
\end{equation}
which is shown in Fig.\,\ref{fig:enerentr}\,(a).
Given this, one is able to obtain the entropy per site (in units of the Boltzmann constant, $k_B$) by\,\cite{Raymond07}
\begin{equation}
\label{entro}
    s(\beta,U) = \ln 4 + \beta e(\beta,U) - \int_{0}^{\beta} e(\beta',U) d\beta',
\end{equation}
which is shown in Fig.\,\ref{fig:enerentr}\,(b). 
As expected, we obtain $s(T \to \infty) \equiv \ln 4$ for all $U/t$, while it decreases and goes toward zero when $T$ is reduced.
However, the way $s(T) \to 0$ depends on the value of $U/t$.
For instance, for $U/t=3$, the entropy approaches zero in close similarity to the non-interacting case, while for $U/t=7$ it decays slowly at low temperatures, directly affecting the specific heat, as discussed latter.

Interestingly, the entropy curves for different values of the interaction strength cross around $s \approx{\ln 2}$.
This crossing has also been observed for the square \cite{Paiva01,Khatami11,Tang12,Tang13} and honeycomb lattices, \cite{Tang12,Tang13,Paiva05} and is closely connected with the possibility of adiabatical cooling in the system.
That is, for entropies greater than $s \approx \ln 2$, Fig.\,\ref{fig:enerentr}\,(b) shows that increasing $U/t$ (at fixed entropy) pushes the temperature up.
On the other hand, below $s \approx \ln 2$, increasing $U/t$ at fixed entropy actually cools the system.
At this point, we recall that this change of behavior should occur around an energy scale where the entropy of a Fermi liquid state becomes lower then the limit of the Heisenberg model (i.e., of localized spin-$1/2$).
At high temperatures, the latter exhibits weakly interacting spins, which, in turn, leads to a crossing of entropies around $s \approx \ln 2$.

\begin{figure}[t]
\includegraphics[scale=0.28]{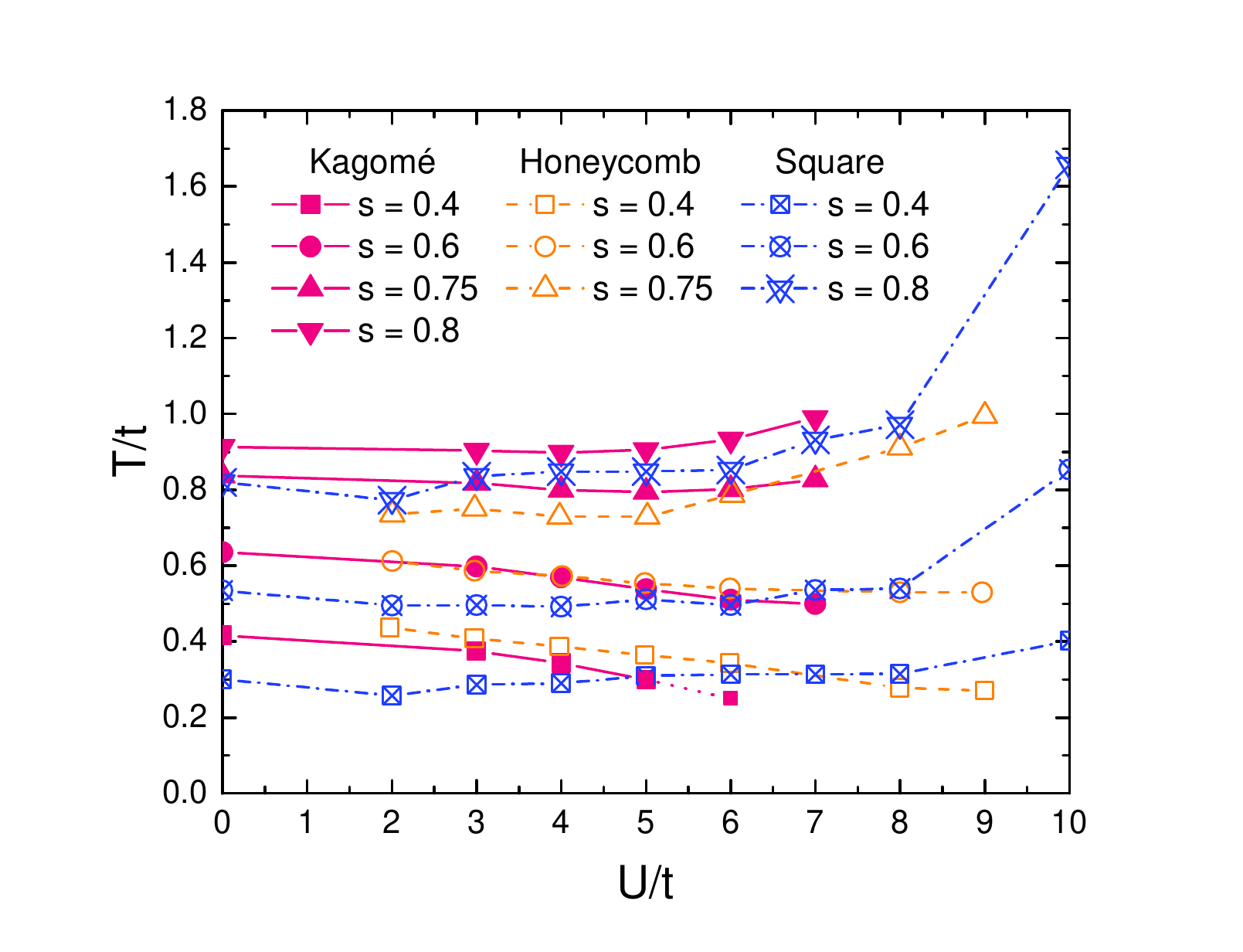} 
\caption{(Color online) Isentropic curves as a function fo $U/t$ for the kagome (closed symbols), honeycomb (open symbols) and square (crossed symbols) lattices.}
\label{fig:isentcomp}
\end{figure}

The adiabatic cooling/heating of the systems is shown in Fig.\,\ref{fig:isentcomp}, which is constructed by selecting fixed values of entropy in Fig.\,\ref{fig:enerentr}\,(b), and gathering the temperature for each $U/t$ value.
A comparison between the isentropic curves on the square,\cite{Paiva10} honeycomb\cite{Tang12} and kagome lattices is also presented and shows that, for low values of $s$, adiabatic cooling on the kagome lattice is at least as effective as in the honeycomb lattice.
Further evidence of it is present below when investigating the behavior of the double occupancy at lower temperatures.

\begin{figure}[t]
\includegraphics[scale=0.24]{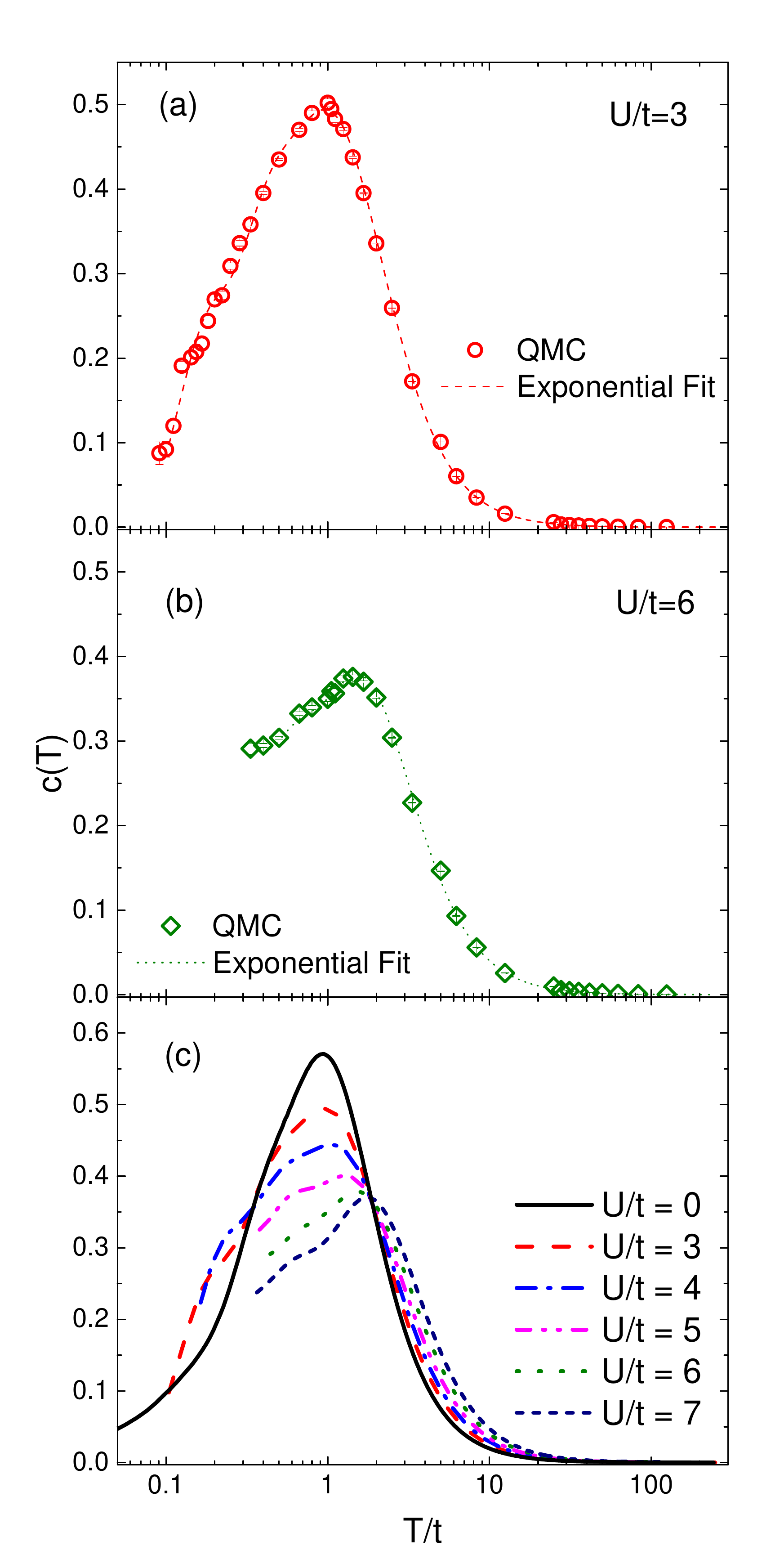} 
\caption{(Color online) The specific heat as function of temperature for different values of \textit{U/t}. The curves in panels (a) and (b) display a comparison between the numerical differentiation of the raw DQMC energy values and the differentiation of the exponential fit performed for $U/t = 3$ and $U/t = 6$, respectively. (c) The specific heat obtained from the exponential fit. When not shown, error bars are smaller than symbol size.}
\label{fig:spheat}
\end{figure}

Figures \ref{fig:spheat}\,(a) and (b) display the behavior of the specific heat,
\begin{equation}
\label{caloresp}
    c(T) = \frac{1}{N} \frac{d \langle \mathcal{H} \rangle}{dT},
\end{equation}
for $U/t=3$ and 6, respectively.
The data points correspond to the differentiation of the raw QMC results in Fig.\,\ref{fig:enerentr}\,(a), while the solid lines are obtained by differentiating a nonlinear fit of the energy; we use an exponential fit of the energy by the function $e_{fit}(T) = a_{0} + \sum_{n=1}^{M} a_{n} \exp(-\beta n \Delta)$, with a cut-off in $M=6$.
Figure \ref{fig:spheat}\,(c) presents the specific heat from the exponential fit, for all values of $U/t$ examined. 
For the noninteracting case, one notices the occurrence of a single peak around $T/t \approx 1$, which is pushed up to higher temperatures as $U/t$ increases.
Such a high-temperature broad peak is due to single-particle excitations, and is closely related to the formation of local moments\,\cite{Paiva01}.
When the temperature is reduced a soft shoulder appears for all $U > 0$, but without a large second peak, as displayed in Fig.\,\ref{fig:spheat}\,(c).
The occurrence of such low-temperature peak is usually due to low-lying collective spin-wave excitations, with the double-peak structure indicating strong spin-spin correlations\,\cite{Paiva01}.
Interestingly, by employing a cellular DMFT approach, Udagawa and Motome\,\cite{Udagawa10} obtained a small second peak for the specific heat at low temperatures, which they suggested is related to spin chirality degrees of freedom. 
In view of these issues, we further investigate the magnetic properties of the Hubbard model in the kagome lattice in the next subsection.

\subsection{Magnetic properties}
\begin{figure}[t]
\includegraphics[scale=0.24]{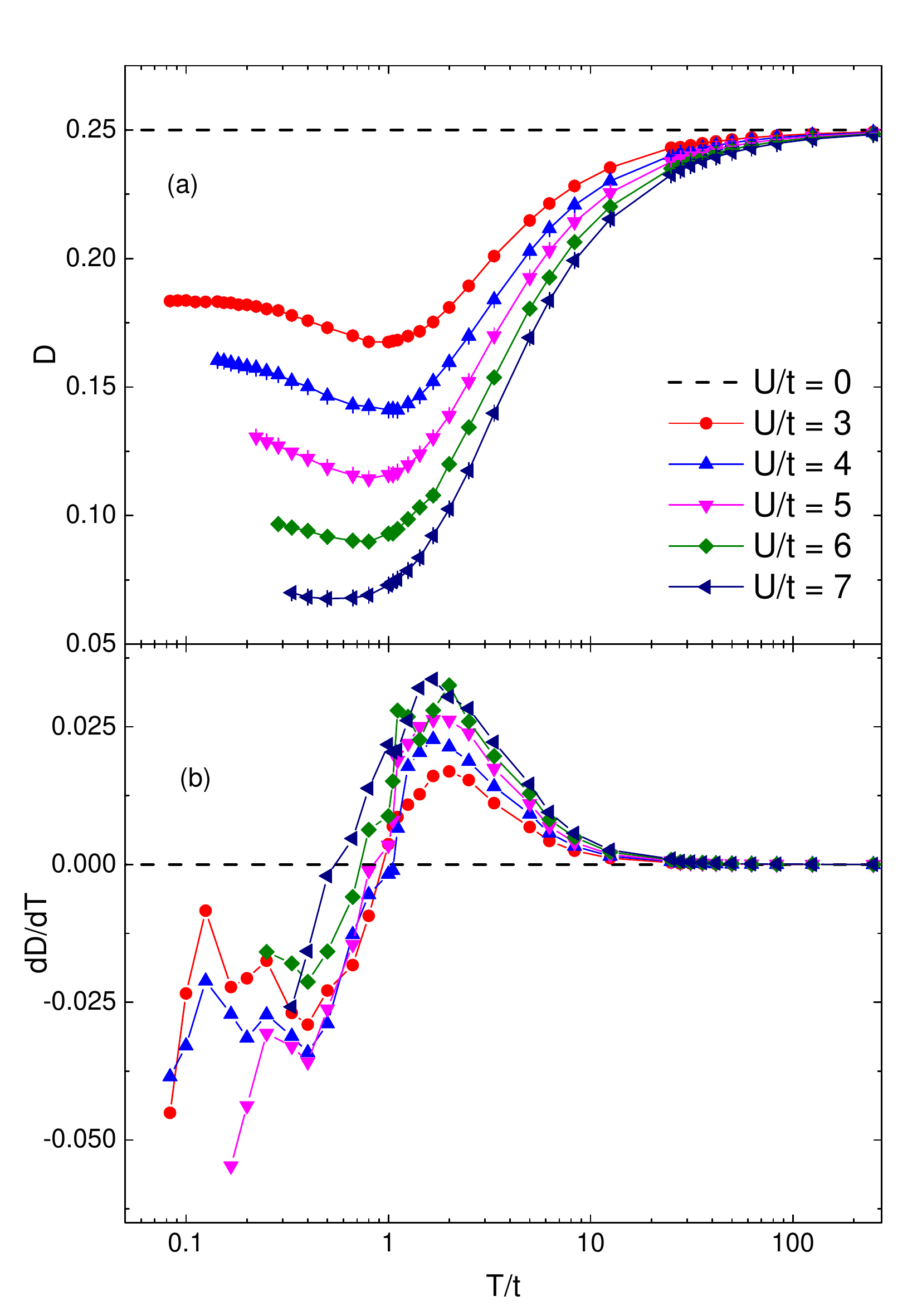} 
\caption{(Color online) (a) Double occupancy and (b) its derivative as functions of the temperature for different \textit{U/t}. The derivatives were smoothed through usual methods. When not shown, error bars are smaller than symbol size.}
\label{fig:docc}
\end{figure}

Now, we turn our attention to the magnetic properties of the system, starting our analysis with the double occupancy,
\begin{equation}\label{eq:docc}
    D = \frac{1}{3 L^2} \bigg\langle \sum_{i, \alpha} n_{i \uparrow} n_{i \downarrow} \bigg\rangle.
\end{equation}
The double occupation and the local moment are connected by $D = \frac{1}{2} \left[ \langle n \rangle - \langle m^2 \rangle  \right] $, 
therefore for fixed $n$, increasing the local moment reduces the double occupancy.  
Figure \ref{fig:docc}\,(a) displays $D$ as a function of temperature, for different values of $U/t$, where one can notice that $D$ has a sharp decrease for $1 < T/t < 10$ with a minimum (for all values of \textit{U}) around $T/t \lesssim 1$.
This minimum suggests a competition between localization and delocalization of the fermions.
As depicted in Fig.\,\ref{fig:docc}\,(b), when $T \to 0$ we find $\frac{\partial D}{\partial T} < 0$ for all $U/t$; a feature consistent with a metallic behavior.
For $U/t=7$, this minimum is shallow, while having a large local moment, which suggests an insulating or a bad metallic behavior.
At this point it is worth mentioning that metallic or insulating behavior is closely connected with of adiabatic cooling features.
Indeed, since $\left( \frac{\partial D}{\partial T} \right)_{N, U} = - \left( \frac{\partial S}{\partial U} \right)_{N, T}$, when $\left( \frac{\partial S}{\partial U} \right)_{N, T} > 0$ -- in the adiabatic cooling region --, then $\left( \frac{\partial D}{\partial T} \right)_{N, U} < 0$, which is consistent with a Fermi liquid at lower temperatures.

\begin{figure}[t]
\includegraphics[scale=0.24]{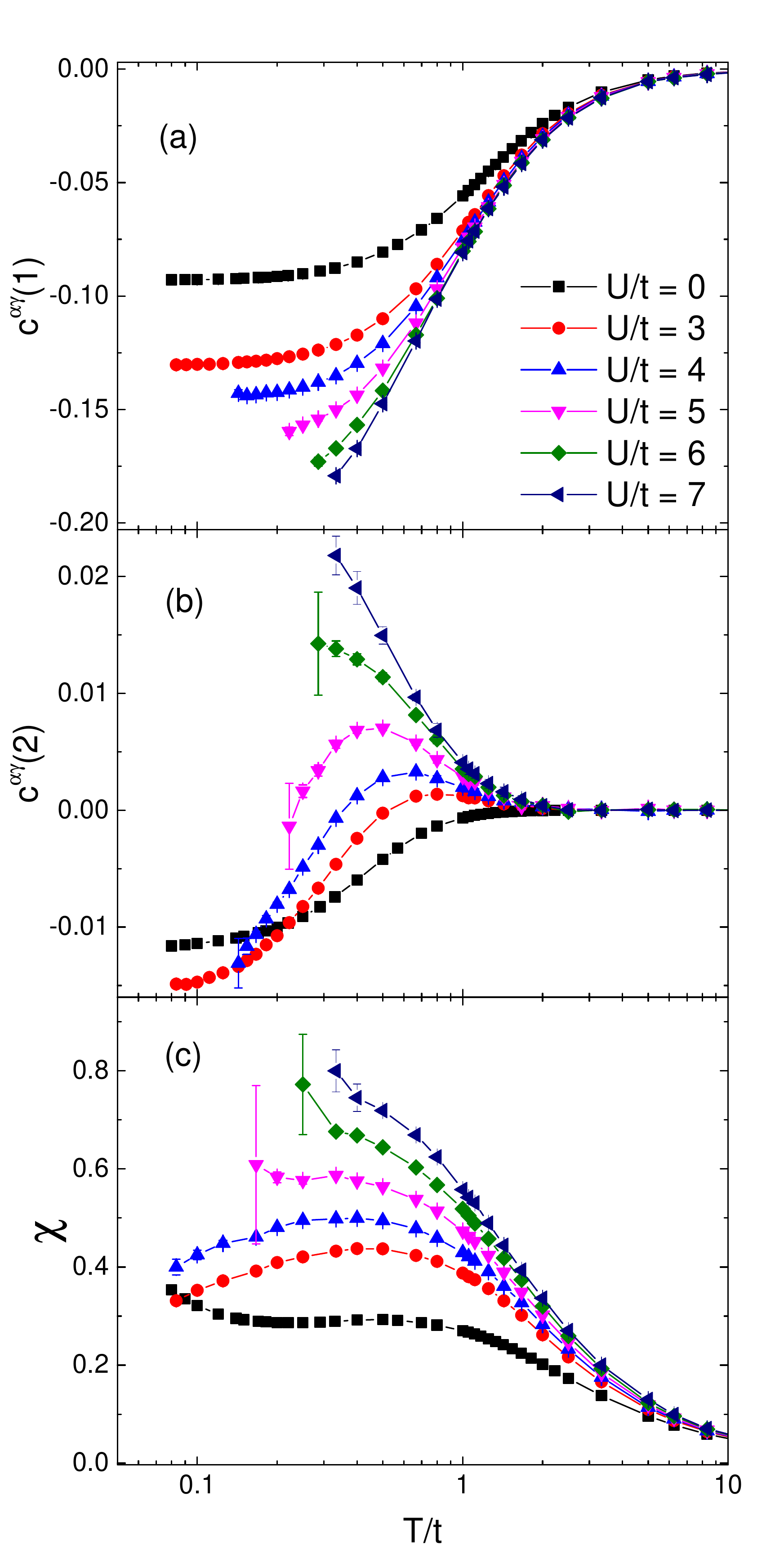}
\caption{(Color online) The spin-spin correlation between (a) nearest neighbors, (b) next-nearest neighbors and (c) the homogeneous magnetic susceptibility as functions of temperature for different values of \textit{U/t}. When not shown, error bars are smaller than symbol size.}
\label{fig:spin_cor_sus}
\end{figure}

We proceed probing nonlocal spin-spin correlation functions
\begin{equation}
\label{spinspin}
    c^{\alpha \gamma} (\textbf{i} - \textbf{j}) = \frac{1}{3} \langle \vec{S}_{\textbf{i},\alpha} \cdot \vec{S}_{\textbf{j},\gamma} \rangle,
\end{equation}
with $\vec{S}_{\textbf{i},\alpha}=(S^{x}_{\mathbf{i},\alpha}, S^{y}_{\mathbf{i},\alpha}, S^{z}_{\mathbf{i},\alpha})$ being the spin operator of a fermion in given unit cell $\mathbf{i}$, and site index $\alpha=A$, \textit{B}, and \textit{C}.
We first explore the nearest neighbors case $c(1)$, when $|\textbf{i} - \textbf{j}| = a$ (lattice parameter), i.e., the spin-spin correlations between the pairs of sites that form a triangle (see Fig. \ref{fig:kagome}). Fig.\,\ref{fig:spin_cor_sus}\,(a), shows $c(1)$ averaged over all the combinations of near-neighbor pairs in the lattice.
$c(1)$ is negative, and increases in magnitude as ${U/t}$ increases, i.e.~there are strong spin-spin correlations along the sides of the triangle.

\begin{figure}[t]
\includegraphics[scale=0.24]{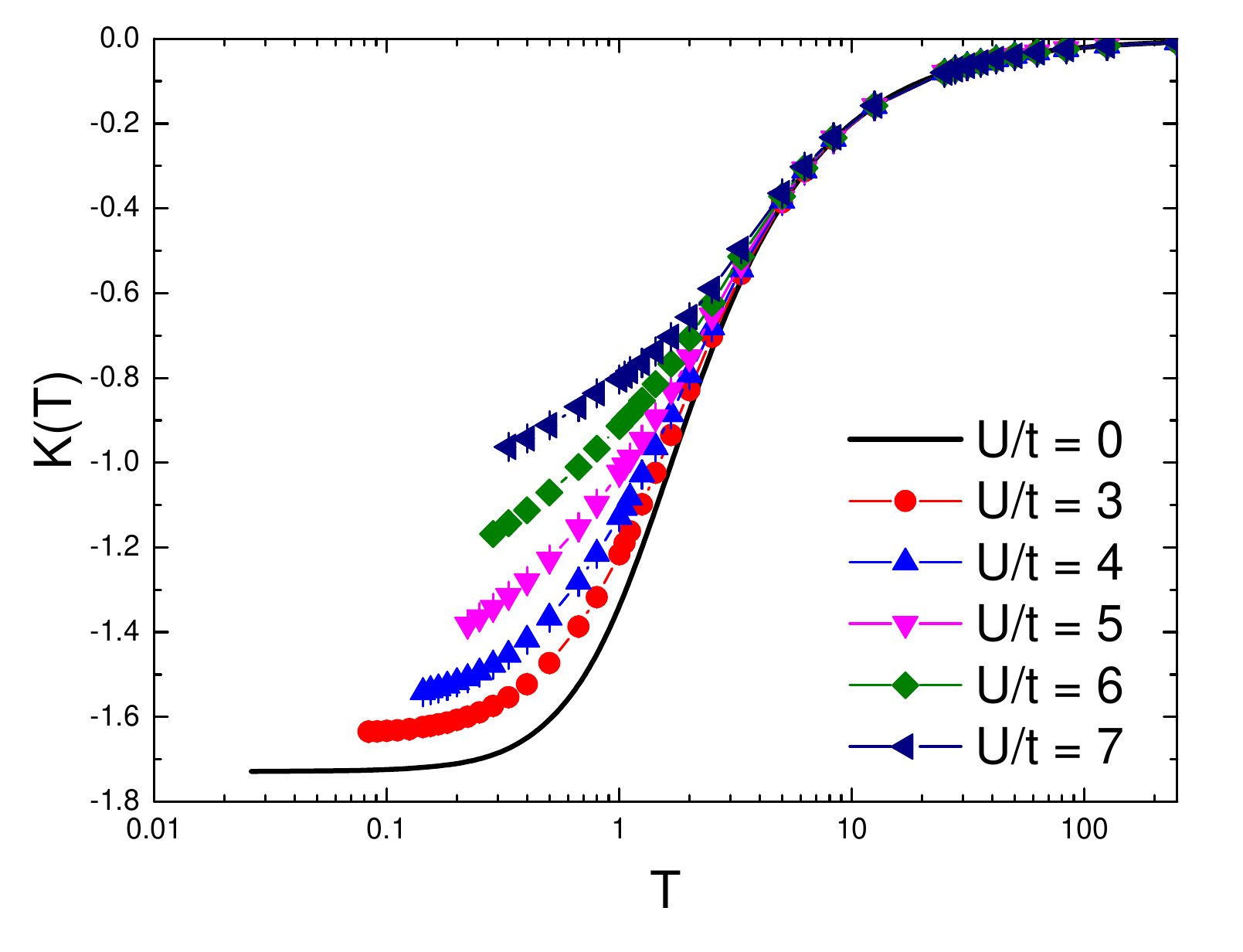} 
\caption{(Color online) The average of the kinetic energy term as a function of the temperature for different values of $U/t$. When not shown, error bars are smaller than symbol size.}
\label{fig:kinetic}
\end{figure}

On the other hand, spin correlations for longer distances are suppressed, as displayed in Fig.\,\ref{fig:spin_cor_sus}\,(b) for next-nearest neighbors  [$c(2)$, when $|\textbf{i} - \textbf{j}| = 2a$], presenting values one order of magnitude smaller than those of $c(1)$.
For $U/t \leq 4$, and at low temperatures, i.e.~below the energy scale for local moment formation ($T/t \lesssim 2$), $c(2)$ is reduced to values very close to those of the non-interacting case, in line with a non-magnetic state.
Interestingly, for larger values of $U$, as for $U/t \geq 5$, $c(2)$ exhibits ferromagnetic correlations at high temperatures, and the effects of frustration (antiferromagnetic correlations) only set in for lower $T/t$.
Finally, Fig.\,\ref{fig:spin_cor_sus}\,(c) displays the homogeneous susceptibility as a function of temperature.
For $U/t \lesssim 5$, $\chi(T)$ has a response similar to the non-interacting case, therefore being consistent with a metallic Pauli paramagnetic state, while for $U/t \gtrsim 6$ it has a monotonically increasing behavior as $T$ decreases (within the temperature range we have analyzed).
However, this increase in $\chi(T)$ with decreasing $T$ for large interaction strengths is not enough to define the nature of the spin excitations. The minus-sign problem prevents us to reach the very low temperatures required to determine if excitations are gapped or gapless.

\subsection{Transport properties}
Lastly, we now investigate the transport properties of the system starting with the kinetic energy.
Figure \ref{fig:kinetic} displays the kinetic energy per site, $\langle \hat{K} \rangle =  -\frac{t}{N} \big\langle \sum_{\substack{\langle \textbf{i},\textbf{j} \rangle},\sigma} \big( c_{\textbf{i} \sigma}^{\dagger}c_{\textbf{j} \sigma}+ {\rm H.c.} \big) \big\rangle$, as a function of temperature for different values of interaction.
Since fermionic localization is favored in an insulating state, therefore $\langle \hat{K} \rangle$ has to be reduced as a function of temperature as $U$ increases, a feature noticed in Fig.\,\ref{fig:kinetic}.
However, within the temperature scale we investigated, we did not find $ \frac{\partial}{\partial T} \langle \hat{K} \rangle < 0$, therefore, different from the analysis of the potential energy, the behavior of the kinetic energy cannot provide clues on the emergence of an insulating phase.

\begin{figure}[t]
\includegraphics[scale=0.24]{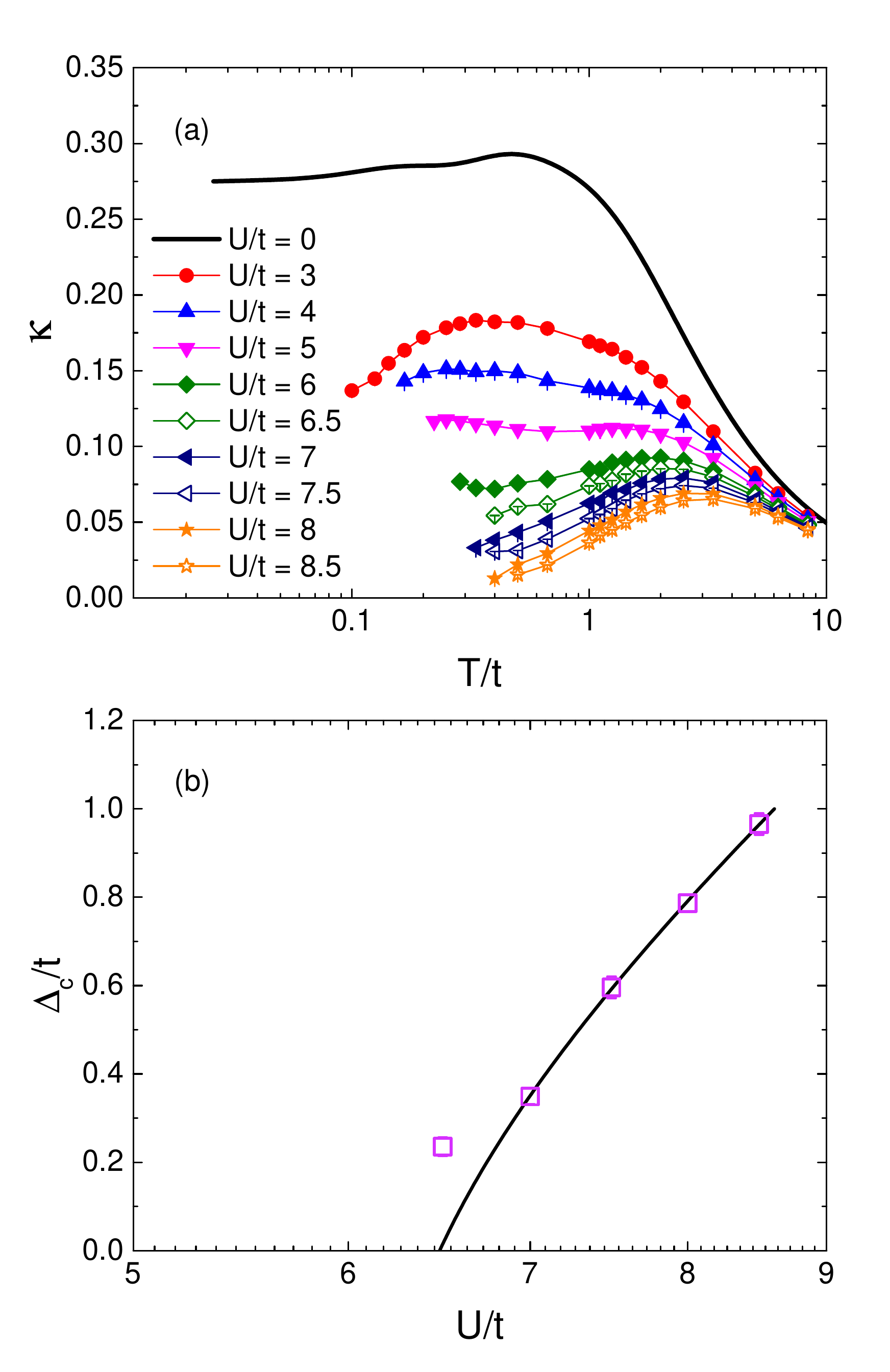} 
\caption{(Color online) (a) The electronic compressibility as a function of the temperature, for different values of $U/t$. (b) The charge gap, obtained through the compressibility data, as a function of $U/t$. When not shown, error bars are smaller than symbol size.}
\label{fig:compress}
\end{figure}

In view of this, other quantities should be investigated to identify the metal-to-insulator transition.
Among them, we proceed to examine the fermionic compressibility,
\begin{equation}
\label{compressi}
    \kappa = \frac{1}{n^{2}} \frac{\partial n}{\partial\mu},
\end{equation}
with $n = \frac{1}{N} \big\langle  \sum_{i, \sigma} n_{i \sigma} \big\rangle $,
displayed in Fig.\,\ref{fig:compress}\,(a).
Notice that, due to the rotation symmetry, the three orbitals are equivalent, as well as their individual compressibilities.
When dealing with any degree of anisotropy (of hopping or interaction), it is possible that some orbitals could be insulating while others exhibit metallic features, as an orbital-selective Mott phase. However, this is beyond the scope of this work.

For a metallic phase, $\kappa$ assumes finite vales, as presented in Fig.\,\ref{fig:compress}\,(a) for the non-interacting case ($U=0$).
For the interacting case, in particular for $U/t \lesssim 5$, the system becomes less compressible, but the trend of $\kappa$ is still consistent with a metallic phase.
Otherwise, for an insulator state, a single-particle gap at the Fermi level of the density-of-states (DOS) is formed, leading to a plateau in $n(\mu)$ as the temperature is reduced, i.e.~to $\kappa \to 0$.
This behavior can be seen for $U/t \gtrsim 6$, where $\kappa$ has a maximum in a high-temperature scale, followed by an exponential suppression at lower temperatures, as depicted in Fig.\,\ref{fig:compress}\,(a).
In other words, there is a clear evidence to a metal-to-insulator transition from the behavior of the compressibility.

We further investigate this change in the compressibility for $U/t \gtrsim 6$ by recalling that, within an insulating state, $\kappa \propto \exp\big(\frac{-\Delta_{c}}{k_{B}T}\big)$, with $\Delta_{c}$ being the charge/single-particle gap ($k_{B}\equiv 1$).
Then, we obtain $\Delta_{c}$ by an exponential fit of $\kappa$ for $U/t \geq 6.5$, as displayed in Fig.\,\ref{fig:compress}\,(b).
Assuming a second-order phase transition, and performing an extrapolation by a polynomial or a power law function for $U/t \geq 7.0$, we obtain the critical point at $U_{c}/t = 6.5 \pm 0.1$, as shown by the black solid line\,\footnote{Notice that the exponential fitting is not adequate when $\Delta_{c} \ll T$, which may lead to misleading results. Therefore, it is safe to disregard the point $U/t = 6.5$ when analysing the extrapolation.}.

At this point, some remarks are required.
First, as the charge gap is formed at high temperature for $U/t \geq 7$, then we expect that our analysis for $\Delta_{c}$ in Fig.\,\ref{fig:compress}\,(b) has little finite-size effects.
Second, we have to recall that, in some circumstances, $\kappa$ may exhibit unconventional behavior in multi-orbital systems, with the charge gap opening only for a few orbitals, while others remain metallic.
However, such an orbital selective Mott transition does not occur in the Hubbard model on the kagome lattice; we have verified that the all orbitals are metallic or insulators.

\begin{figure}[t]
\includegraphics[scale=0.24]{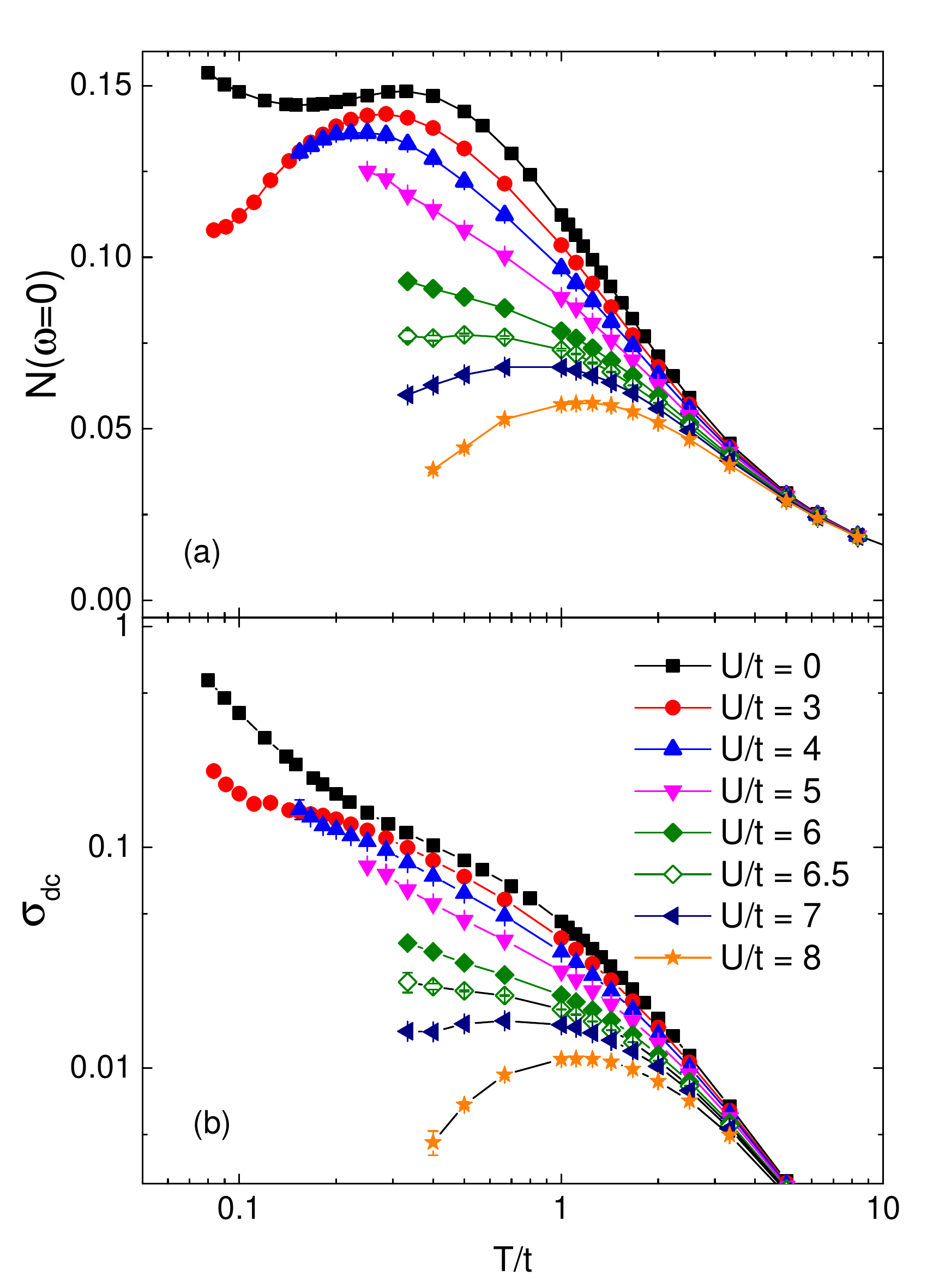} 
\caption{(Color online) (a) The density-of-states at Fermi level, and (b) the dc conductivity as functions of the temperature, for different values of $U/t$. When not shown, error bars are smaller than symbol size.}
\label{fig:sigmadc}
\end{figure}

The previous analysis of $\kappa$ indirectly points out to a suppression of the spectral weight around the Fermi level.
Given this, it is important to \textit{directly} probe the density-of-states (DOS) as a complementary study.
In order to avoid complex methodologies for numerical analytical continuations, here we examine the DOS only at the Fermi level, which is obtained through\,\cite{Trivedi95}
\begin{equation}
\label{eq:densos}
N(\omega = 0) \approx \frac{\beta}{\pi} G(|\mathbf{i}-\mathbf{j}| = 0,\; \tau = \beta/2).
\end{equation}
Figure \ref{fig:sigmadc}\,(a) displays $N(\omega = 0)$ as a function of temperature for different values of $U/t$.
Notice that $N(\omega = 0)$ exhibits a finite value for $U/t \lesssim 5$ at low-temperatures, consistent with a metallic state.
By contrast, for $U/t \gtrsim 6$, the trend of the DOS has a significant change, being reduced exponentially, as expected for an insulator.
In particular, within the range of temperatures examined, the change in behavior occurs at $U/t \approx 6.5$, in very good agreement with the results for the compressibility.

\begin{figure}[t]
\includegraphics[scale=0.15]{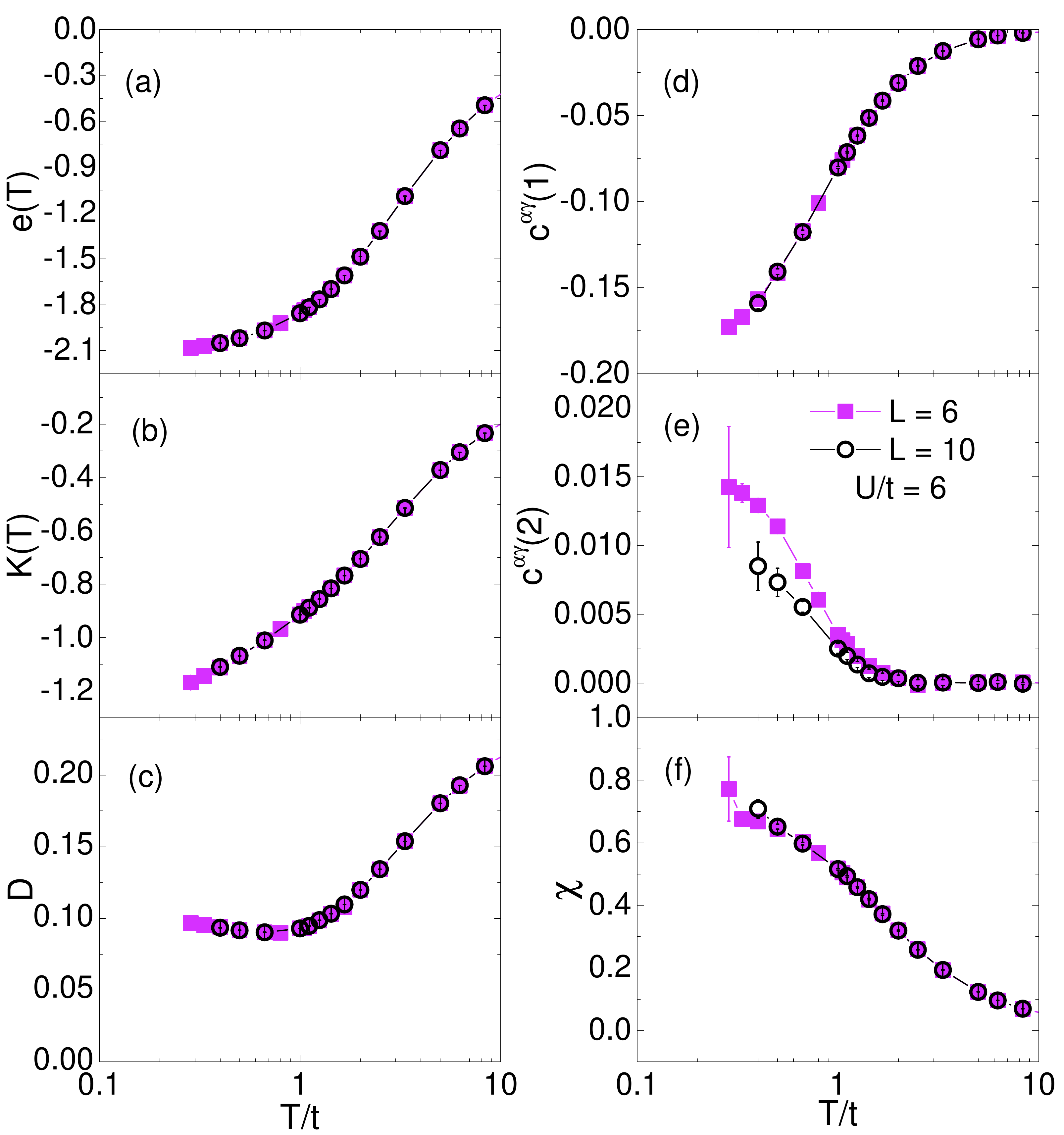} 
\caption{(Color online) Comparison between results obtained for $6 \times 6$ (filled squares) and $10 \times 10$ (open circles) lattices for (a) the total energy, (b) the kinetic energy, (c) the double occupancy, (d) nearest and (e) next-nearest-neighbors spin-spin correlation functions, and (f) homogeneous susceptibility as functions of the temperature, at $U/t=6$. When not shown, error bars are smaller than symbol size.}
\label{fig:size_comp}
\end{figure}

Finally, as a further evidence of a metal-to-insulator transition we examine the dc conductivity,
\begin{equation}\label{eq:sigma_dc}
\sigma_{dc} = \frac{\beta^2}{\pi} \Lambda_{xx}(\mathbf{q=0}, \tau = \beta/2),
\end{equation}
in which
\begin{equation}
\Lambda_{xx}(\mathbf{q}, \tau ) = 
\langle j_{x}(\mathbf{q}, \tau) j_{x}(-\mathbf{q}, 0)  \rangle,
\end{equation}
with $ j_{x}(\mathbf{q}, \tau) $ being the Fourier transform of the unequal-time current-current correlation functions
\begin{equation}
j_x(\mathbf{i},\tau)=\mathrm{e}^{\tau\mathcal{H}}
  \left[
        it\sum_\sigma
            \left(c_{\mathbf{i}+\mathbf{x}\sigma}^\dagger 
                  c_{\mathbf{i}\sigma}^{\phantom{\dagger}}
                  - 
                  c_{\mathbf{i}\sigma}^\dagger  
                  c_{\mathbf{i}+\mathbf{x}\sigma}^{\phantom{\dagger}}
            \right)
  \right]
\mathrm{e}^{-\tau\mathcal{H}};
\label{jx}
\end{equation}
see, e.g., Refs.\onlinecite{Trivedi96,Denteneer99,Mondaini12}.
Figure \ref{fig:sigmadc}\,(b) exhibits the results for $\sigma_{\rm dc}$ as a function of temperature for different values of $U/t$. 
Similarly to the previous analyses, a metallic behavior, i.e.~$\partial \sigma_{\rm dc} / \partial T < 0$, occurs only for $U/t \lesssim 6.5$.
For interaction strength larger than that, the behavior is consistent with an insulator.

In summary, the analyses of the compressibility, the DOS at the Fermi level, and the dc conductivity provide strong evidence for a metal-to-insulator transition at $U_{c}/t = 6.5 \pm 0.5$.


\subsection{Finite-size effects}

In this subsection, we investigate finite-size effects.
To this end, we perform simulations for a $N = 10 \times 10$ lattice (i.e., with 300 sites), fixing $U/t=6$, while varying temperature.
Figure \ref{fig:size_comp} displays (a) the total energy, (b) the kinetic energy, (c) the double occupancy, and (d) nearest neighbors spin-spin correlation functions for this system size, comparing it with our previous results for $N = 6 \times 6$.
Notice that, for these quantities, finite-size effects may be disregarded within the range of temperature examined.
Indeed, short-ranged quantities must suffer much less from finite-size effects than long-ranged quantities, as structure factors or susceptibilities.
Concerning the latter, NNN spin correlation functions and the homogeneous magnetic susceptibility exhibit a small dependence on the lattice size at low temperatures, as shown in Figs.\,\ref{fig:size_comp}\,(e) and (f), respectively.
However, these minor dependencies do not affect the main results discussed in the previous subsections.



\begin{figure}[t]
\includegraphics[scale=0.30]{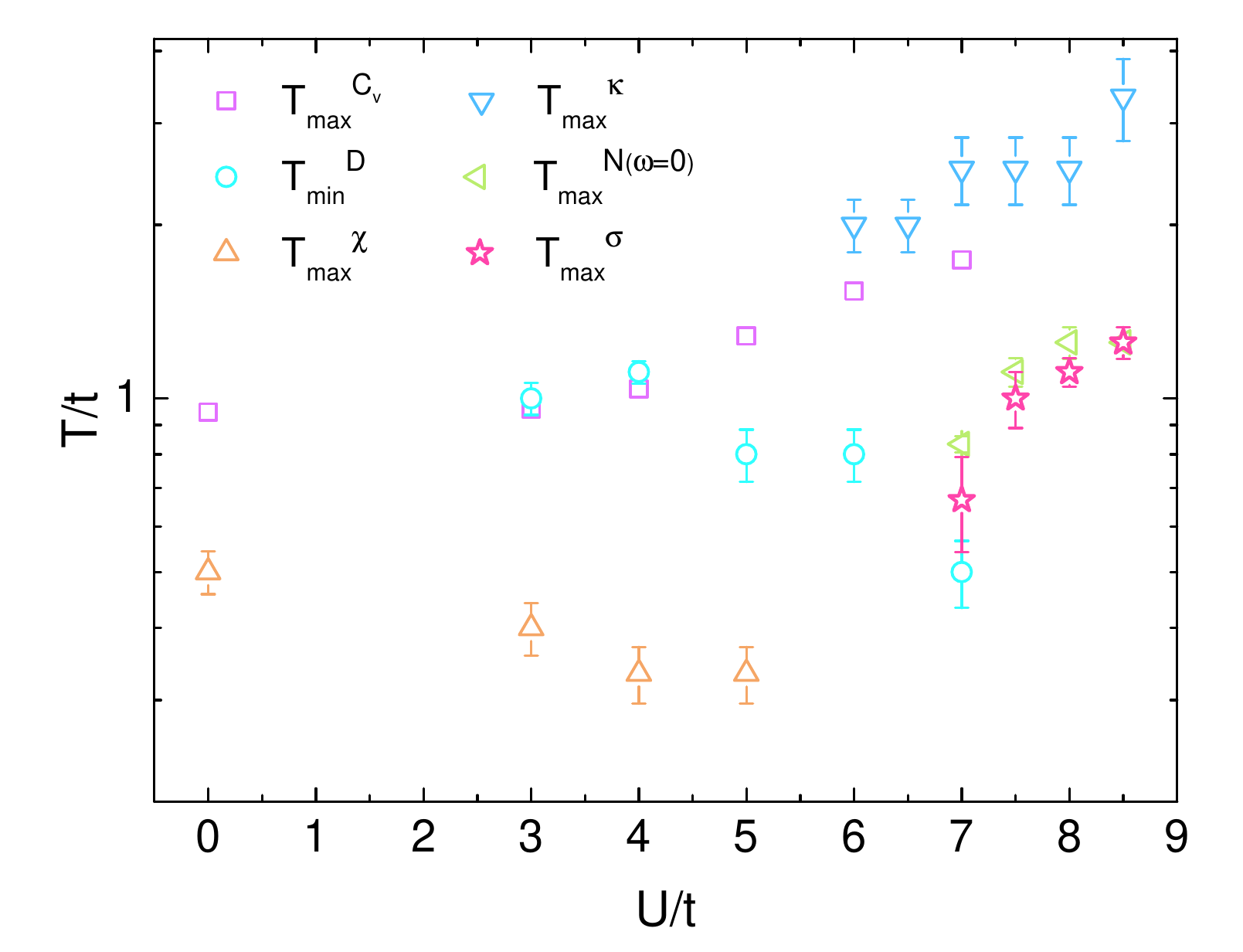} 
\caption{(Color online) Different energy scales for the Hubbard model on the kagome lattice, for several values of $U/t$.}
\label{fig:temps}
\end{figure}

\section{Conclusions}
\label{sec:conc}
In this work, we have investigated thermodynamic, magnetic and transport properties of the repulsive Hubbard model on the kagome lattice through unbiased DQMC simulations.
For the thermodynamic properties, we examined the entropy for different interaction strengths, and the behavior of the isentropic curves as a function of $U/t$.
We have found that adiabatical cooling is possible for entropies smaller than $s \approx \ln 2$. 
In addition, we examined the specific heat:
in contrast to what is seen in the Hubbard model at the square or honeycomb lattices, the low temperature peak seems to be suppressed in the kagome lattice for all $U/t$.
This suggests the absence of collective spin-wave excitations as temperature is reduced, i.e.~the absence of magnetic long-range order in the ground state.

In view of this, we investigated the spin-spin correlation functions, in particular the local moment, nearest (NN), and next-nearest neighbors (NNN).
We obtained well-formed local moments, with strong short-range NN correlations functions, while the NNN (and farther) ones are strongly suppressed, further evidencing the absence of magnetic long-range order.
Despite this, the homogeneous magnetic susceptibility $\chi(T)$ still increases as temperature is reduced, being enhanced for larger values of $U/t$.
However, we are not able to conclude whether the nature of the spin excitations is gapped or gapless.
Furthermore, it is worth  mentioning that identifying whether a spin liquid state emerges or not is challenging, and beyond the scope of this work.

Finally, we probed the metal-to-insulator transition.
In particular, the behavior of the compressibility provides a clear distinction between metallic and insulating states.
Therefore, we investigated the behavior $\kappa(T)$ for different values of $U/t$, being able to identify the critical point around $U_{c}/t \approx 6.5$.
As complementary analyses, we also examined the DOS at the Fermi level, as well as the current-current correlation functions, leading to results in line with those from the compressibility.
Together, these analyses provide clear evidence for a Mott transition at $U_{c}/t = 6.5 \pm 0.5$.

In summary, our work presents a detailed finite-temperature analyses for the Hubbard model at the kagome lattice, allowing us to provide different energy scales of the system. 
To this end, Fig.\,\ref{fig:temps} presents [i] the minima of the double occupancy $T_{\rm min}^{D}$, as well as the high-temperature maxima for [ii] the specific heat peak $T_{\rm max}^{c}$, [iii] the magnetic susceptibility $T_{\rm max}^{\chi}$, [iv] the compressibility $T_{\rm max}^{\kappa}$, [v] the DOS at Fermi level $T_{\rm max}^{N}$, and [vi] the dc conductivity $T_{\rm max}^{\sigma}$.
Together, Figs.\,\ref{fig:isentcomp} and \ref{fig:temps} provide a broad description of the model, that can be relevant to future cold atom experiments.


\section*{ACKNOWLEDGMENTS}

We are grateful to J.P.~de Lima for his contributions in the initial stage of this work, and to E.C.~Andrade for illuminating discussions
and suggestions.
Financial support from the Brazilian Agencies CAPES, CNPq,  FAPERJ, and Instituto Nacional de Ciência e Tecnologia de Informação Quântica (INCT-IQ) is gratefully acknowledged.
N.C.C.~acknowledges financial support from CNPq, grant number 313065/2021-7. 



\bibliography{kagome.bib}
\end{document}